\documentclass[aps,onecolumn,prl,showpacs,amsmath,superscriptaddress]{revtex4-2}
\pdfoutput=1
\usepackage{epsfig,graphicx}
\usepackage{bookmark}
\usepackage{booktabs}
\usepackage{physics,amssymb,amsmath,xcolor}
\usepackage{subcaption}
\begin{document}

\title{Protective Measurement - a new quantum measurement paradigm: detailed description of the first realisation}

\author{Enrico Rebufello}
\author{Fabrizio Piacentini}
\author{Alessio Avella}
\author{Marco Gramegna}
\author{Giorgio Brida}
\affiliation{INRIM - Istituto Nazionale di Ricerca Metrologica, strada delle Cacce, 91 10135 Torino, Italy.}

\author{Rudi Lussana}
\author{Federica Villa}
\author{Alberto Tosi}
\affiliation{Politecnico di Milano, Dipartimento di Elettronica, Informazione e Bioingegneria, Piazza Leonardo da Vinci 32, 20133 Milano, Italy.}

\author{Eliahu Cohen}
\affiliation{Faculty of Engineering and the Institute of Nanotechnology and Advanced Materials, Bar Ilan University, Ramat Gan 5290002, Israel.}

\author{Lev Vaidman}
\affiliation{Raymond and Beverly Sackler School of Physics and Astronomy, Tel-Aviv University, Tel-Aviv 6997801, Israel.}

\author{Ivo Pietro Degiovanni}
\author{Marco Genovese}
\affiliation{INRIM - Istituto Nazionale di Ricerca Metrologica, strada delle Cacce, 91 10135 Torino, Italy.}
\affiliation{INFN, sede di Torino, via P. Giuria 1, 10125 Torino, Italy.}

\date{\today}

\begin{abstract}
We present a detailed description of the experiment realising for the first time a protective measurement, a novel measurement protocol which combines weak interactions with a ``protection mechanism'' preserving the measured state coherence during the whole measurement process.
Furthermore, protective measurement allows finding the expectation value of an observable, i.e. an inherently statistical quantity, by measuring a single particle, without the need of any statistics.
This peculiar property, in sharp contrast with the framework of traditional (projective) quantum measurement, might constitute a groundbreaking advance for several quantum technology related fields.
\end{abstract}

\maketitle


\section{Introduction}
Measurement in quantum mechanics is usually described through projective measurements, represented by a projector onto physical states within a given Hilbert space.
This kind of measurement induces the wavefunction collapse onto a specific eigenstate of the observable, corresponding to the observed eigenvalue.
In contrast with this measurement paradigm, in \emph{weak measurements}, introduced by Aharonov, Albert and Vaidman \cite{AAV}, the coupling between the pointer and the quantum state is weak, introducing only a partial decoherence of the wavefunction, at the price of acquiring only partial information about the state. Examples of weak-coupling-based schemes are measurement of weak values \cite{AAV,eli} and protective measurements (PM) \cite{AHARONOV199337,Meaning,vaidmanrev1,vaidmanrev2}.

In general, a common property for all quantum measurements is that the measurement procedure is invasive, inducing unavoidable decoherence in the initial state of the system.
Even in weak value measurements \cite{AAV,eli,FerrieCoin,MundarainWV,vaidmancontr,Ex1,Ex2,Ex4,WhiteLight,pusey2,qust,SinclairToy,LundNat,GoggPNAS,A4,stjoint,Ex5,avella,seqth,sequential, thekk,kimseq,vallone1,vallone2,vallone3,valencia1,valencia2,scirep,elinat,georgiev,wwpan,ywcho,shik1,shik2,shik3,shik4},
the coupling between the system and the measuring device causes a small perturbation to the system state.
In contrast with other quantum measurement paradigms, a PM is able to preserve the coherence of the quantum state during the whole measurement process, thanks to a \textit{protection mechanism} or, alternatively, via the adiabatic theorem \cite{AHARONOV199337}.
This difference with respect to traditional measurement protocols allows PM to extract the expectation value of an observable (so far considered as an inherently statistical quantity, only obtainable by means of repeated measurement of an ensemble of identically-prepared systems) by measuring a single quantum system.

Hence, PM is a novel measurement paradigm presenting significant elements of interest both as a tool for quantum metrology and for understanding the very foundations of quantum measurement and, more generally, of quantum mechanics itself, e.g. the possibility of measuring a stationary wavefunction $\ket{\psi}$ \cite{AHARONOV199337}.
For this reason, PMs add significant elements to the debate about the ontic or epistemic nature of the wavefunction, a highly debated topic in the scientific community \cite{Meaning,vaidmanrev1,vaidmanrev2,wavefunction,Genovese:2010:1936-6612:249,doi:10.1142/S0217979213450124,Pusey2012,Rovelli,Unruh,Dickson,d,Dass,Uffink,ShanUffink,Ringbauer2015,Prot4, gao,Khrennikov,pan}.\\

In this work, we extensively illustrate the scheme, methodology and obtained results related to the first experimental implementation of PM \cite{Piacentini2017}, demonstrating its capability to preserve the system state coherence and, at the same time, extract the expectation value of an observable even from a single measurement event.

\subsection{Theoretical framework}
In the framework of quantum mechanics, given a quantum state $\ket{\psi}$, we define the quantum expectation value of an observable $A=\sum_{i}a_i|\phi_i\rangle\langle\phi_i|$ (with $\sum_{i}|\phi_i\rangle\langle\phi_i|=\mathbb{I}$) as the average of its eigenvalues $a_i$ weighted on their respective probabilities $p_i$:
\begin{equation}
\ev*{A}=\mel{\psi}{A}{\psi}=\sum_i p_i a_i \;\;\;\;\;\;\;\; \big(p_i=|\langle\psi|\phi_i\rangle|^2\big)
\end{equation}
Similarly to its classical counterpart, $\ev*{A}$ is understood as a statistical property.\\

PM can be modeled as a standard von Neumann measurement \cite{vonneumann} in which we couple the observable of interest $A$ with a pointer $P$ with a long and adiabatic interaction instead of the usual instantaneous one.
Such interaction is mediated by the coupling $g(t)$, which allows us to write the interaction Hamiltonian as
\begin{equation}\label{H}
  \mathcal{H}_{int} = g(t) A \otimes P
\end{equation}
where the interaction intensity is $g(t)=g/T$ for a time interval $T$ and smoothly goes to zero before and after.
If the coupling $g(t)$ is smooth enough, we obtain the adiabatic limit, in which the state of the system $\ket{\psi}$ does not change, thanks to the protection.
In the limit $T \to \infty$ and for bounded $P$, one has $\mathcal{H}_{int} \to 0$ while the state $\ket{\psi}$ remains unchanged thanks to the preserving action of the protection mechanism. We can calculate, then, the shift of the energy of the eigenstate via first-order perturbation theory \cite{AHARONOV199337}:
\begin{equation}
  \delta E \simeq \ev*{\mathcal{H}_{int}} = g\frac{\ev*{A}P}{T}
\end{equation}
from which we can calculate the time evolution $U$ associated to $ \mathcal{H}_{int}$ in the limit $T \to \infty$:
\begin{equation}
  U \simeq \exp \left( -\frac{i}{\hbar} g \ev*{A}P \right)
\end{equation}
resulting in a pointer wavefunction shift proportional to the expectation value $\ev*{A}$.\\
{A second protection scheme involves the so-called \textit{active} protection, based on the quantum Zeno effect \cite{zeno}, which consists of a series of repeated projections onto the initial state \cite{itano,kwiatz,kk1,kk2,kk3,fischer,pasc1,pasc2,smerzi1,smerzi2,signoles,ghera1,ghera2,ghera3,ghera4,kur1,kur2,kur3,vedral,qze-aze} during the interaction described by the Hamiltonian in Eq. (\ref{H}), but in the non-adiabatic limit.
In our experiment, we will focus on this quantum Zeno-type protection \cite{vaidmanrev2}.}\\

From a quantum informational perspective, our experiment corresponds to a protocol in which Alice produces a quantum state, which she transmits to Bob together with the proper protection, implemented by Bob as a black box when realizing the protective measurement.\\

The PM framework can be formalized with an equivalent description consisting of a series of $K$ instantaneous weak interactions, described by a (weak) coupling constant $g=\int_{t=0}^{T}g(t)dt$.
Between two subsequent interactions, the active protection occurs, induced by the projector $\Pi_{\psi} = \dyad{\psi}$.
It is straightforward to show that, in the weak interaction approximation ($g \ll 1$), each of the $K$ weak interaction/protection blocks evolves the system in the following way:
\begin{equation}
  \ket{\psi} \! \! \! \ev*{U}{\psi} \otimes \ket{\varphi(x)} = \ket{\psi} \! \! \! \ev*{\exp \left( - \frac{i}{\hbar} g A \otimes P \right)}{\psi} \otimes \ket{\varphi(x)} \approx \ket{\psi} \otimes \ket{\varphi(x-g\ev*{\hat{A}})}
\end{equation}\label{PM1}
being $\ket{\varphi(x)}$ the pointer initial wavefunction.\\
From Eq. (\ref{PM1}), one can see how the PM induces a shift in the meter wavefunction, which is directly proportional to the expectation value $\ev*{A}$.
PMs, then, allow us to directly estimate the expectation value $\ev*{A}$ for each single particle undergoing them, in sharp contrast with the concept of $\ev*{A}$ being only a statistical quantity.\\
\section{Experimental implementation}
In our experiment, we implement two different methods to measure the expectation value of the polarization operator $A=\dyad{H}-\dyad{V}$ (with $H$ and $V$ being the horizontal and vertical polarization, respectively): the aforementioned PM, able to estimate $\ev*{A}$ with a single reading of the measuring device, and a traditional projective measurement, in which the expectation value is extracted from the statistics obtained from repeated measurements on an ensemble of identically-prepared particles.
Both measurements can be described as a von Neumann protocol in which we couple the polarization of an incoming photon, prepared in the linearly polarized state $\ket{\psi} = \cos (\theta) \ket{H} + \sin (\theta) \ket{V}$, with its transverse momentum $\mathbf{P}$:
\begin{equation}
  U = \exp \left( - \frac{i}{\hbar} g \Pi_H \otimes \mathbf{P} \right)
\end{equation}
where $\Pi_H = \dyad{H}$ is the projector onto the $H$ polarization.
This interaction causes a shift of the horizontally-polarised component of the wavefunction along an axis orthogonal to the photon propagation direction.
This is mathematically equivalent to a von Neumann coupling of strength $g/2$ between the polarization $A$ and the momentum $\mathbf{P}$, so from here we will consider a rescaling of our system in order to describe the latter scenario.\\

\begin{figure}
\centering
\includegraphics[width=0.8\textwidth]{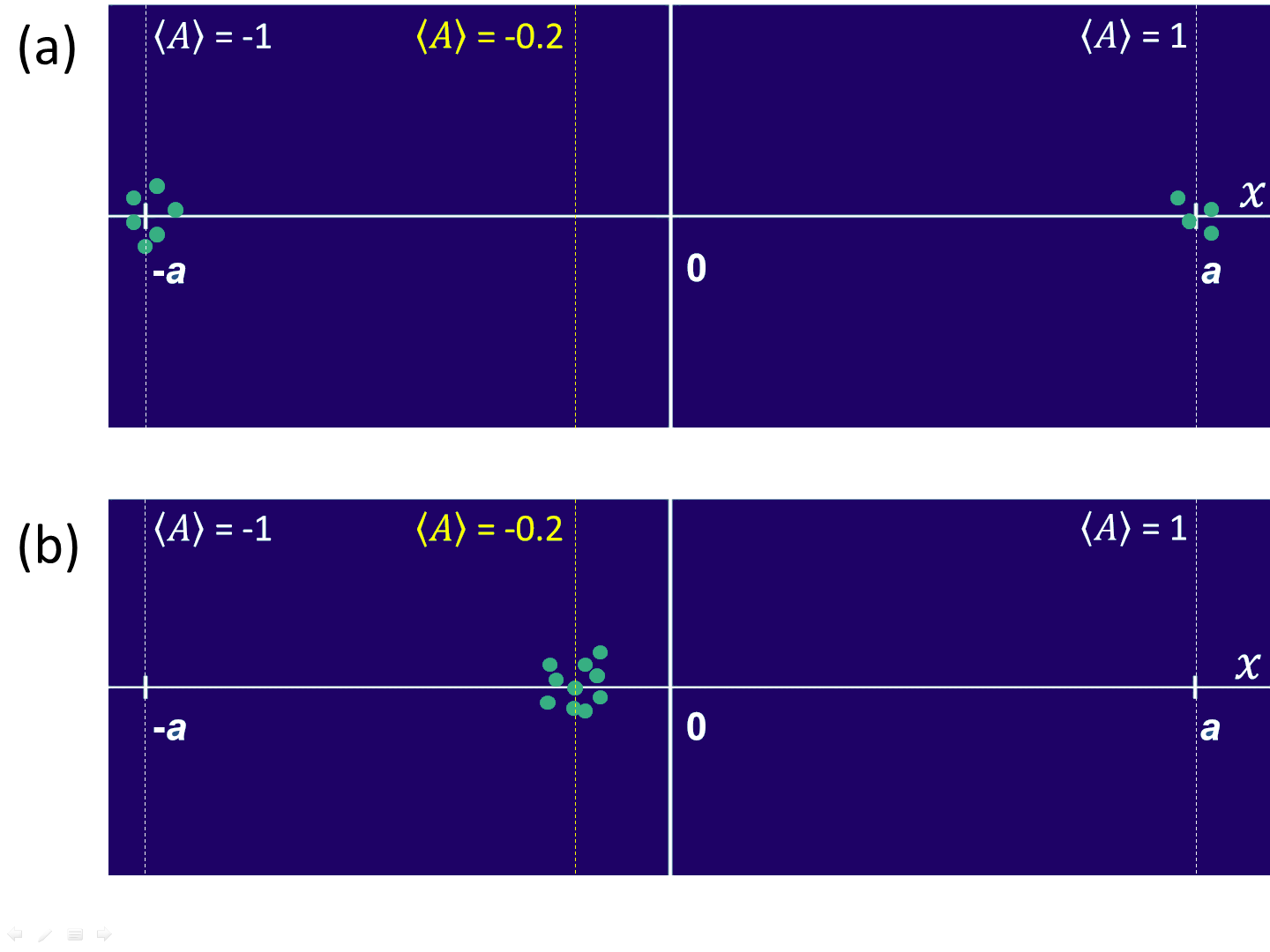}
\caption{Theoretical framework (ideal pictorial representation). (\textbf{a}) Projective measurement: the two polarization components are completely separated, with the single photons impinging on two regions of the detector corresponding to the polarization operator eigenvaulues $A=\pm1$. The expectation value $\ev*{A}$ is evaluated as the weighted average of the events, following Eq. (\ref{A_PJ}). (\textbf{b}) Protective measurements: all the photons fall in the same region, centered in a position proportional to the polarization expectation value (see Eq. (\ref{A_PM})).}
\label{theo}
\end{figure}

The initial spatial wavefunction of the photon is described by a normal distribution:
\begin{equation}
\phi_0(x)=\braket{x}{\phi_0}=\frac{1}{\sqrt[4]{2\pi\sigma^2}}\exp \left( -\frac{(x-x_0)^2}{4\sigma^2} \right),
\end{equation}
centred at $x_0$ and with standard deviation $\sigma$.\\
For strong interactions (i.e. $g \gg 1$), the two polarization components will be completely separated (Fig. \ref{theo}a).
Hence, the expectation value can be evaluated as:
\begin{equation}
  \ev*{A} = \frac{N_H - N_V}{N_H + N_V}
  \label{A_PJ}
\end{equation}
being $N_{H(V)}$ the number of count events obtained for the polarization $H$($V$).
This is the case of projective measurement \cite{Gerlach1922}.\\

PMs, instead, in our scheme consist of a series of weak von Neumann couplings ($g \ll 1$) alternating with a protection mechanism, i.e. a projection $\Pi_\psi = \dyad{\psi}$ onto the initial state $\ket{\psi}$. In this case, the photons will fall in a region not corresponding to any eigenvalue of our polarization observable $A$, but whose position is directly proportional to its expectation value (Fig. \ref{theo}b).
Thus, the expectation value of the polarization can be extracted by the formula
\begin{equation}
  \ev*{A} = \frac{x-x_0}{a}
  \label{A_PM}
\end{equation}
with $x_0 = \frac{x_H+x_V}{2}$ and $a= \frac{x_H-x_V}{2}$, being $x_H$ ($x_V$) the center of the horizontally-(vertically-)polarized photon distribution in the projective measurement framework.

\subsection{Experimental Setup}

\begin{figure}
\centering
\includegraphics[width=0.98\textwidth]{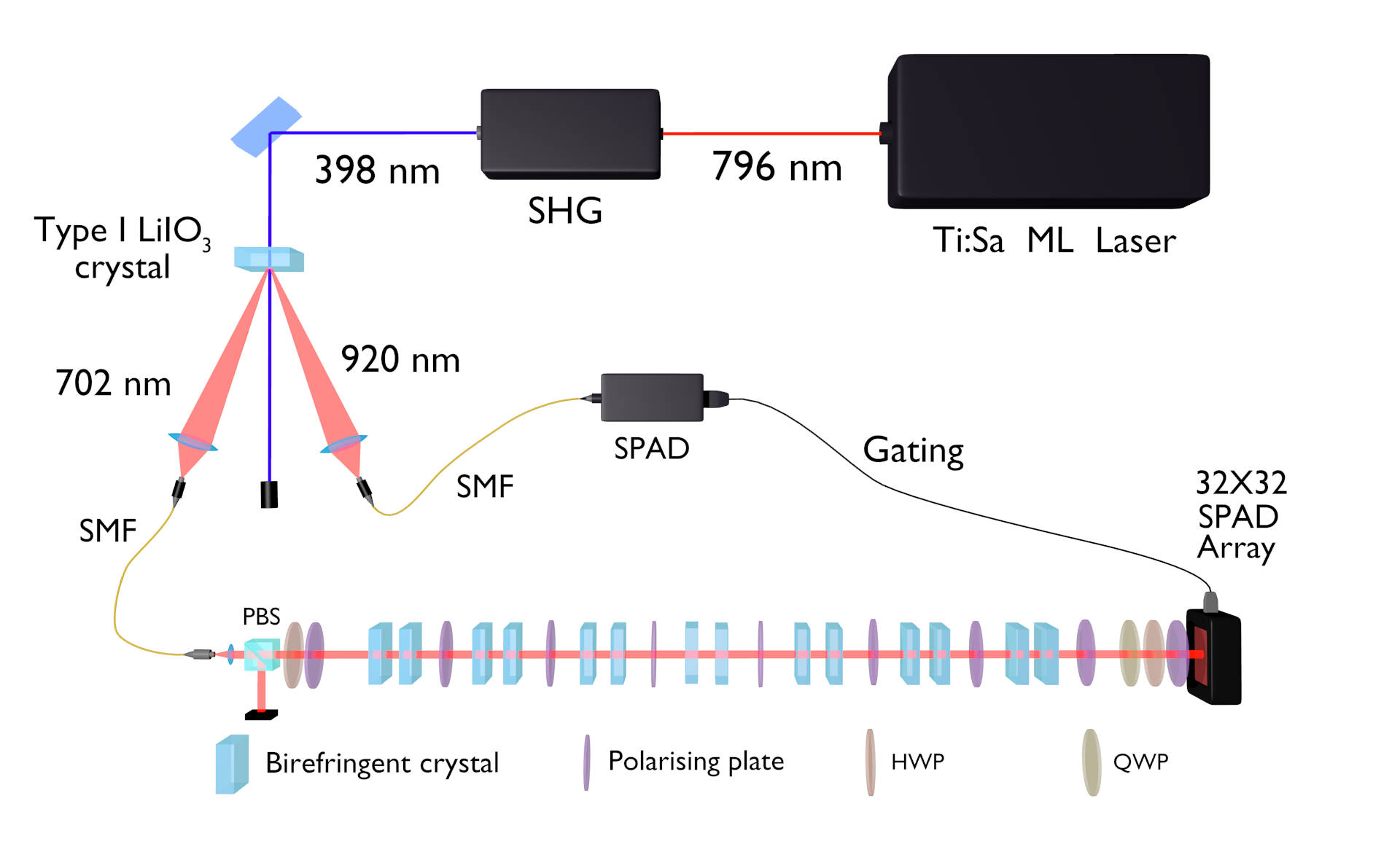}
\caption{Experimental setup. SHG: second harmonic generator; SPAD: single-photon avalanche diode; SMF: single-mode fiber. HWP: half-wave plate. QWP: quarter-wave plate. PBS: polarising beam splitter. Ti:Sa ML laser: titanium-sapphire mode-locked laser.}
\label{setup}
\end{figure}

Our experimental setup (Fig. \ref{setup}) is divided in three parts.
In the first part, single photons are generated by a heralded single-photon source \cite{eis,sinha}.
A mode-locked laser with a second harmonic at 398 nm and a 76 MHz repetition rate pumps a $10\times10\times5$ mm LiIO$_3$ non-linear crystal, producing signal-idler photon pairs by exploiting Type-I spontaneous parametric down-conversion (SPDC).
The generated idler photons (920 nm) are filtered by an interference filter (IF) centred at 920 nm and with a FWHM of 10 nm, and coupled to a single-mode fibre (SMF) which addresses them to a Silicon single-photon avalanche diode (SPAD), heralding the presence of the correlated signal photons (702 nm).
Signal photons are filtered with an IF centred at 702 nm and with a FWHM of 10 nm, fibre-coupled and addressed to the second part of the setup, where the PM takes place.
A Hanbury-Brown and Twiss interferometer allowed us to estimate the quality of our single-photon emission by evaluating the $\alpha$ parameter \cite{alpha,alphaivo}, obtaining a value of $\alpha = 0.13(1)$ without any background or dark count subtraction, testifying the high quality of our heralded single-photon source.\\
In the second part, the protective and projective measurements are performed.
The signal photon produced in the previous stage is decoupled and collimated in a Gaussian spatial mode over a 2 m length.
Then, it is initialized (pre-selected) in the polarization state $\ket{\psi}$ by a polarizing beam splitter (PBS) followed by a half-wave plate (HWP).
Finally, in the projective measurement configuration the photon goes through $K=7$ weak interaction units (we chose the number of units $K=7$ from practical considerations approximating the ideal case of large $K$).
Each unit consists of a first 2 mm long birefringent calcite crystal with an extraordinary ($e$) optical axis lying in the $x$ - $z$ plane, having an angle of $\pi/4$ with respect to the $z$ direction, followed by a second 1.1 mm calcite crystal with the optical $e$-axis oriented along the $y$-axis.
The first crystal shifts the horizontally-polarized component of the wavefunction along the transverse direction $x$, while the second one compensates for the phase and time decoherence induced by the first crystal.
The combined effect of all the 7 units allows for complete separation of orthogonal polarizations, reproducing the projective measurement framework.
In the PM scenario, instead, the protection of the quantum state is implemented exploiting the quantum Zeno effect, realized by inserting a polarizing plate after each weak interaction unit.
The polarizing plate realizes the projection $\Pi_\psi = \dyad{\psi}$, projecting the state outgoing the weak von Neumann interaction onto the same polarization of the initial state $\ket{\psi}$, thus cancelling the (tiny) decoherence induced by the birefringent crystals in each weak interaction unit.\\

The final part of the experimental setup is the detection stage: photons are detected by a 2D spatial-resolving single-photon detector prototype, i.e. a two-dimensional array of 32x32 ``smart pixels'', each hosting a SPAD with dedicated front-end electronics \cite{6862847}.
The detection of the idler photon (920 nm) by the Si-SPAD on the heralding arm gates the SPAD array with a 6 ns detection window.
Furthermore, an optional quantum tomography \cite{ALTEPETER2005105,shur} apparatus, comprising a HWP, a quarter-wave plate (QWP) and a polarizer, can be inserted before the SPAD array to reconstruct the density matrix of the single-photon state at the end of each measurement procedure.\\
%
%
%
\section{Results}
We acquired data sets for three different states: the state $\ket{+} = \frac{1}{\sqrt{2}} (\ket{H}+\ket{V})$, which should be subjected to the maximum decoherence, and two intermediate states $\ket{\frac{17}{60}\pi}=\cos(\frac{17}{60}\pi)\ket{H}+\sin(\frac{17}{60}\pi)\ket{V}$ and $\ket{\frac{\pi}{8}}=\cos(\frac{\pi}{8})\ket{H}+\sin(\frac{\pi}{8})\ket{V}$.
Each data set is composed of multiple acquisitions:
\begin{itemize}
    \item An acquisition with only the crystals in the optical path and $\ket{\psi}=\ket{H}$ or $\ket{\psi}=\ket{V}$, which allows us to calibrate the system.
    \item An acquisition without protection (only crystals in the optical path), corresponding to the traditional projective measurement scenario.
    \item An acquisition with both weak interaction and active Zeno-like protection (both birefringent crystals and polarisers in the optical path), realizing the PM.
    \item Two acquisitions, one with only the polarizing plates and one with a free optical path, allowing to complete the system calibration by evaluating and properly subtracting unwanted position biases introduced by crystals and polarizing plates.
\end{itemize}
\subsection{Output state verification}
To immediately highlight the difference between PM and projective (PJ) measurements, we perform a tomographic reconstruction of the states at the end of the measurement process.\\
%
In the PJ case, we expect that the repeated shifts of the horizontal polarization component cause decoherence on $\hat{\rho}_{in}= \dyad{\psi}$, generating a final state $\hat{\rho}_{dec}$.
On the contrary, in the PM case the protection should be able to preserve, in principle, the initial polarization state $\hat{\rho}_{in}$, thus we expect a final state $\hat{\rho}_{prot} = \hat{\rho}_{in}$:
\begin{align}
  \hat{\rho}_{in} &= \begin{pmatrix}
	\cos^2 (\theta) & - \sin(\theta) \cos(\theta)\\
	\sin(\theta) \cos^(\theta) & \sin^2(\theta)
  \end{pmatrix}\\
  \hat{\rho}_{dec} &= \begin{pmatrix}
	\cos^2 (\theta) & - \sin(\theta) \cos(\theta) \exp \left( -\frac{g^{\prime^2}}{(2\sigma)^2} \right)\\
	\sin(\theta) \cos(\theta) \exp \left( -\frac{g^{\prime^2}}{(2\sigma)^2} \right) & \sin^2(\theta)
  \end{pmatrix}
  \label{eq:stati-th-prot}
\end{align}
where $g^\prime = \ev*{x_H}-\ev*{x_V}=11.56(7)$ px (pixels), being $\ev*{x_{H(V)}}$ the average position of photons in the $H(V)$ polarization on the $x$ axis, and $\sigma=4.17(2)$ px is the distribution width obtained by Gaussian fits.\\

\begin{figure}[h]
\centering
\includegraphics[width=0.9\textwidth]{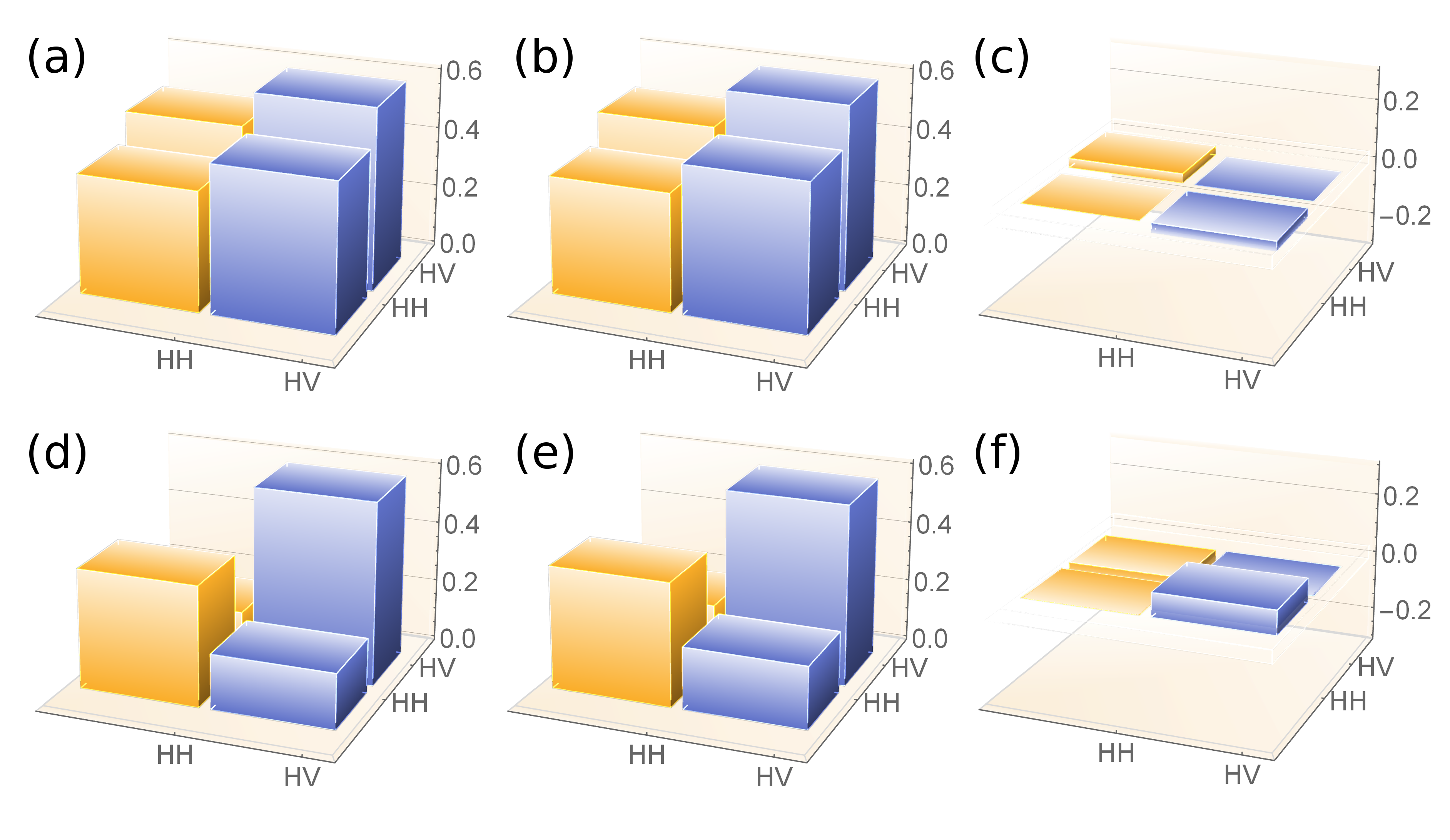}
\caption{Density matrix reconstructions for the state outgoing the measurement process, considering an initial polarization state $\ket{\frac{17}{60}\pi}$: (\textbf{a}) Theoretical real part $\Re [\rho_{in}]$ of the initial state density matrix ($\Im [\rho_{in}]=0$). (\textbf{b}) and (\textbf{c}), respectively: reconstructed real ($\Re [\rho_{PM}^{rec}]$) and imaginary ($\Im [\rho_{PM}^{rec}]$) part of the density matrix of the single-photon state after the protective measurement. (\textbf{d}) theoretically-expected real part $\Re [\rho_{dec}]$ of the density matrix of our state at the end of the projective measurement ($\Im [\rho_{dec}]=0$). (\textbf{e}) and (\textbf{f}), respectively: reconstructed real ($\Re [\rho_{PJ}^{rec}]$) and imaginary ($\Im [\rho_{PJ}^{rec}]$) part of the density matrix of the single-photon state after undergoing the projective measurement.}
\label{tomo}
\end{figure}

We first compute the distance between the reconstructed states $\rho_{PM}^{rec}$ and $\rho_{PJ}^{rec}$, respectively obtained in the protective and projective measurement case (shown in Figure \ref{tomo} for the state $\ket{\frac{17}{60}\pi}$), and their theoretical counterparts $\rho_{in}$ and $\rho_{dec}$, by evaluating the Fidelity $F$ \cite{nielsenchuang,geometry} between them (second and third columns of Table \ref{tableFP}).
The high Fidelities obtained certify the adherence of the experimental results to our model, showing that PM indeed preserves the initial state by the decoherence induced by the birefringent crystals, while this does not happen for projective measurements.\\

Then, we compute the distance between the two reconstructed states (fourth column of Table \ref{tableFP}).
Again, the low fidelities obtained tell us that, without protection, the decoherence induced on the initial state by the $K=7$ unitary interactions makes the final state totally incompatible with the one outgoing the PM procedure (and, obviously, with the initial state itself).\\

Finally, by analysing the purity $\mathcal{P}$ \cite{nielsenchuang} of the reconstructed states (last two columns of Table \ref{tableFP}), we notice that, as expected, the decoherence reduced the purity of the initial state $\rho_{in}$ in the PJ case, while this does not happen in the PM one.
Thus, we have proved the PM protocol ability to preserve the coherence of the initial state during the whole measurement process, a feature in sharp contrast with the traditional quantum measurement paradigms.
%
\begin{table}[h]
  \caption{Comparison between theoretical and reconstructed density matrices. $F(\rho_{PM}^{rec}, \rho_{in})$ and $F(\rho_{PJ}^{rec}, \rho_{dec})$: Fidelities between reconstructed density matrices $\rho_{PM(PJ)}^{rec}$ and their theoretical counterparts $\rho_{in(dec)}$ in the PM (PJ) case. $F(\rho_{PM}^{rec}, \rho_{PJ}^{rec})$: Fidelities between reconstructed protected and unprotected states. $\mathcal{P}(\rho_{PM}^{rec})$ and $\mathcal{P}(\rho_{PJ}^{rec})$: Purities of the reconstructed states in the PM and PJ case, respectively.}
\centering
\begin{tabular}{cccccc}
\toprule
\textbf{State}	& $F(\rho_{PM}^{rec}, \rho_{in})$ & $F(\rho_{PJ}^{rec}, \rho_{dec})$ & $F(\rho_{PM}^{rec}, \rho_{PJ}^{rec})$ & $\mathcal{P}(\rho_{PM}^{rec})$ & $\mathcal{P}(\rho_{PJ}^{rec})$\\
\midrule
$\ket{+}$ & 0.999 & 0.998 & 0.720 & 0.998 & 0.540\\
$\ket{\frac{17}{60}\pi}$ & 0.996 & 0.999 & 0.751 & 0.992 & 0.520\\
$\ket{\frac{\pi}{8}}$ & 0.992 & 0.999 & 0.894 & 0.992 & 0.789\\
\bottomrule
\end{tabular}
\label{tableFP}
\end{table}
\subsection{Expectation values}
Subsequently, in order to test the predictions of PM regarding the possibility of extracting the expectation value $\ev*{A}$ even from a single detection event, we evaluate $\ev*{A}$ with both the PM and PJ methods.\\
\begin{figure}[!p]
\centering
    \begin{subfigure}[b]{0.48\textwidth}
        \centering
        \includegraphics[width=\textwidth]{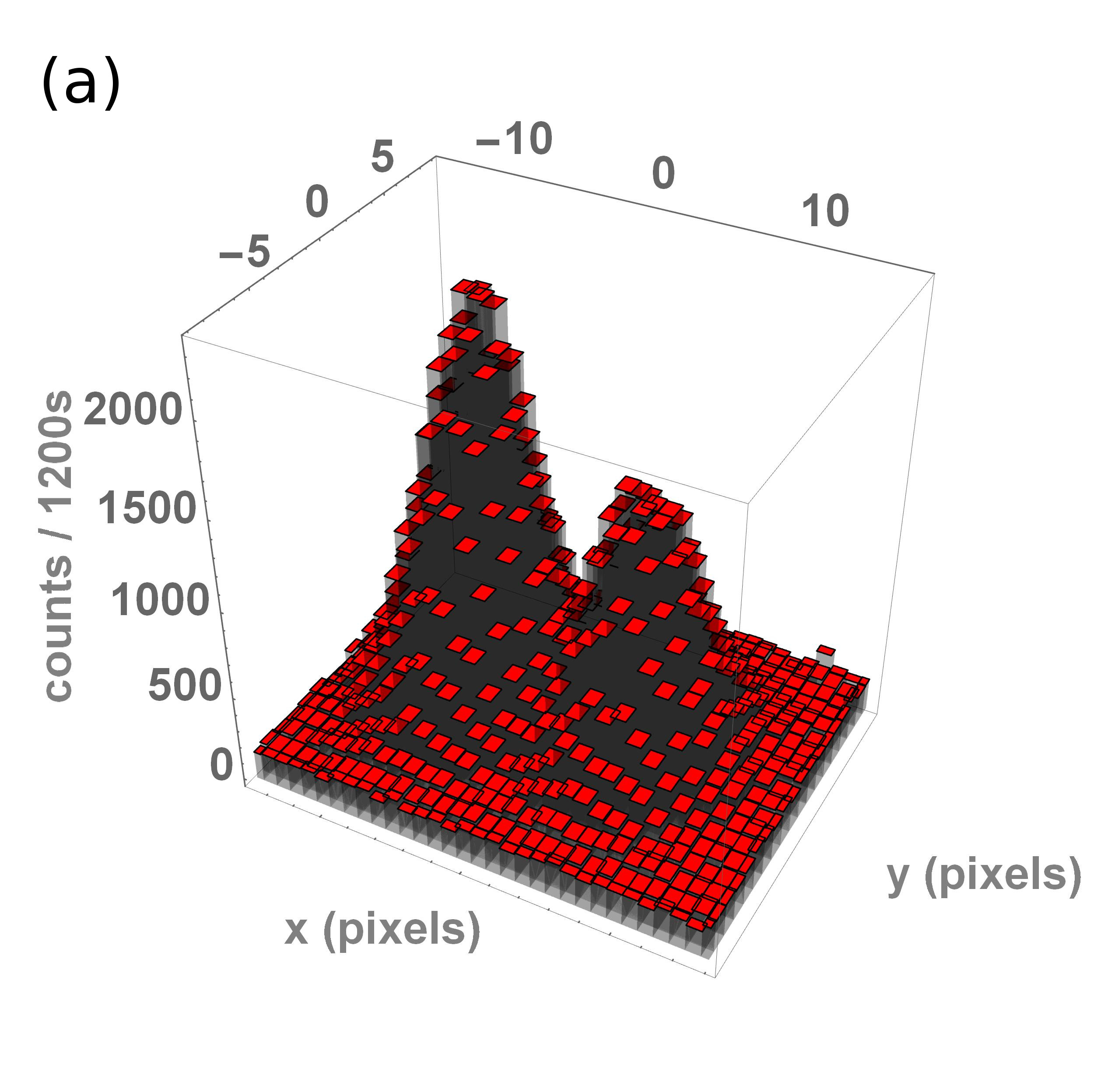}
    \end{subfigure}
    \begin{subfigure}[b]{0.48\textwidth}
        \centering
        \includegraphics[width=\textwidth]{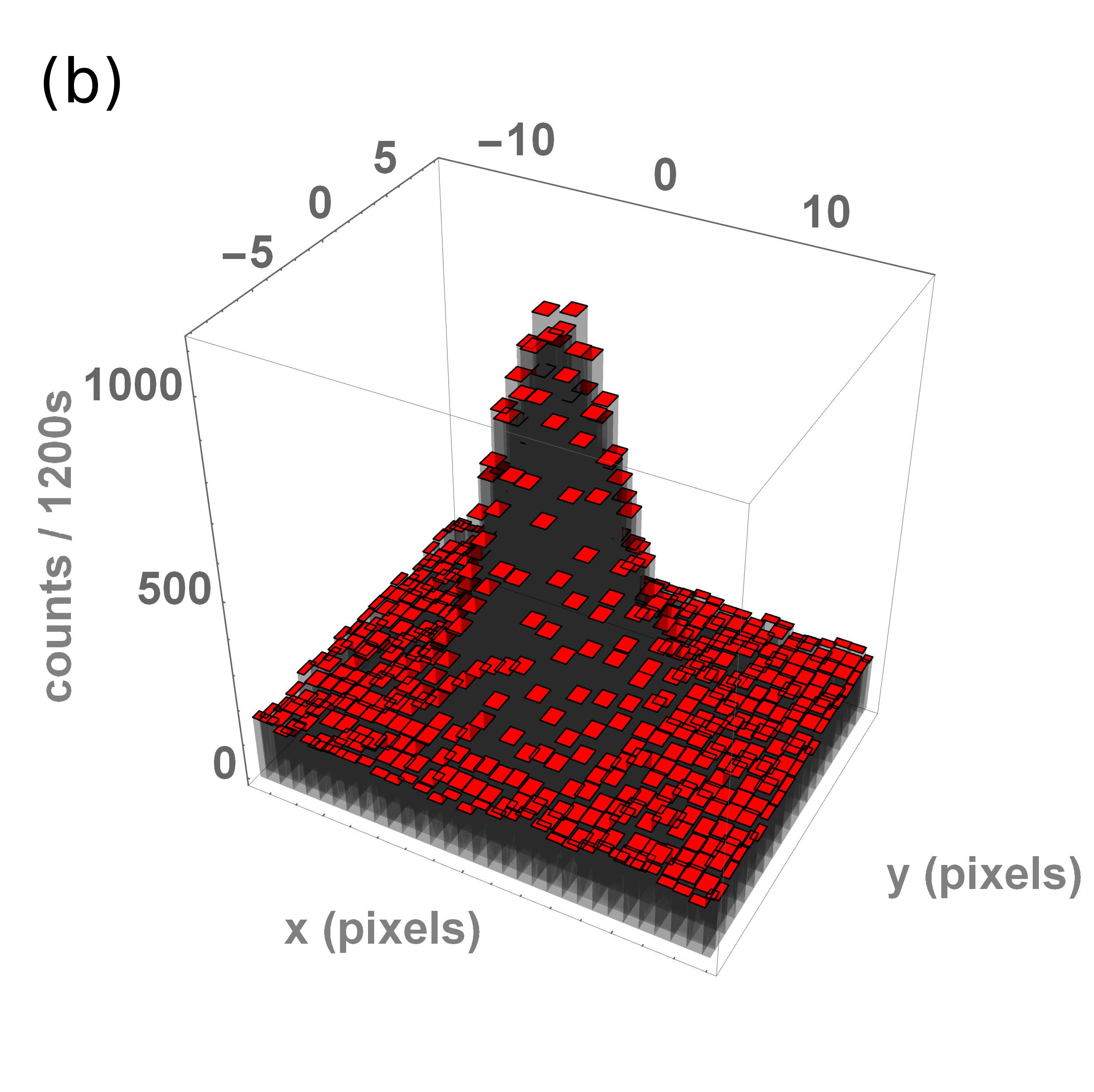}
    \end{subfigure}

    \begin{subfigure}[b]{0.48\textwidth}
        \centering
        \includegraphics[width=\textwidth]{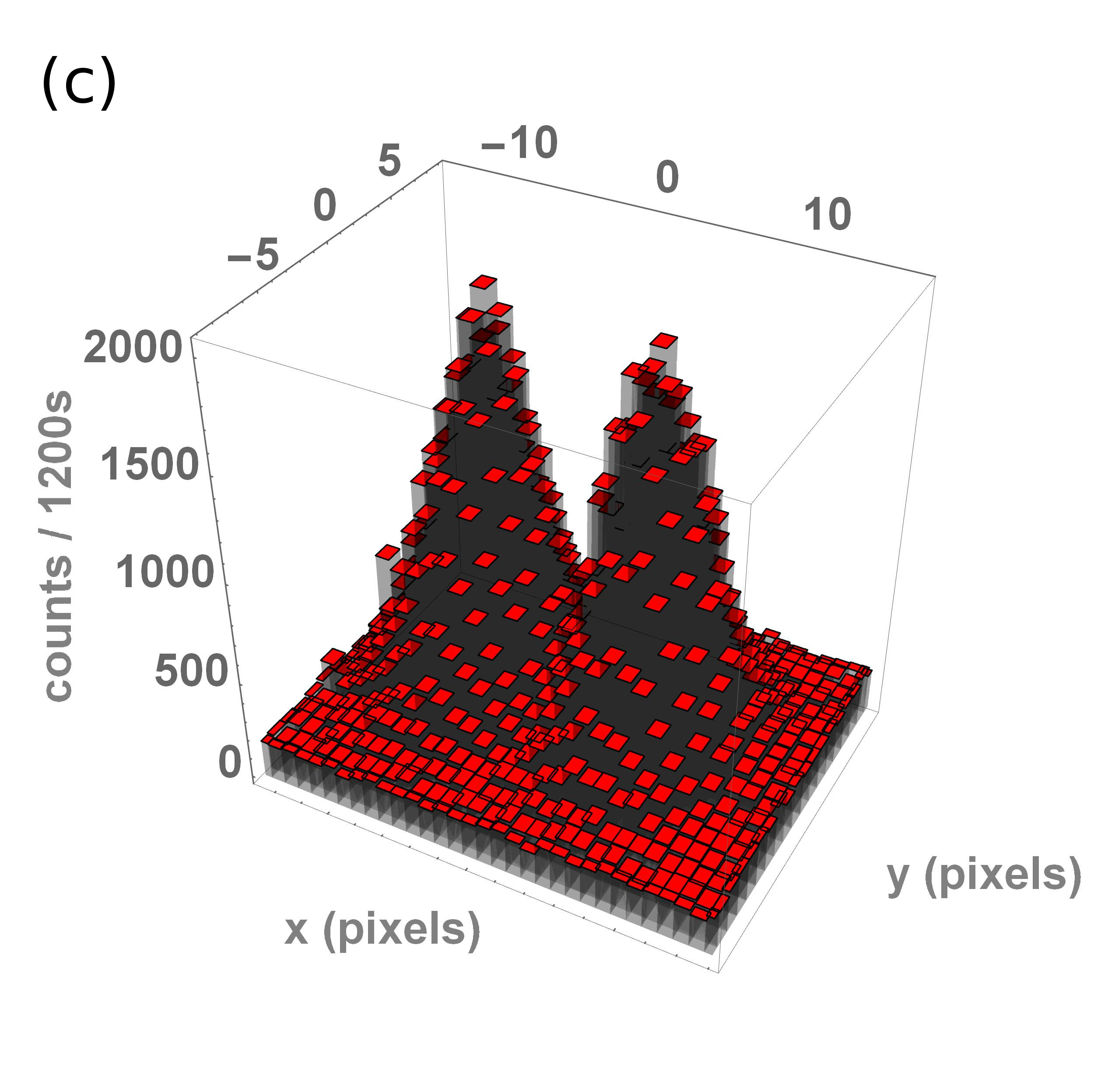}
    \end{subfigure}
    \begin{subfigure}[b]{0.48\textwidth}
        \centering
        \includegraphics[width=\textwidth]{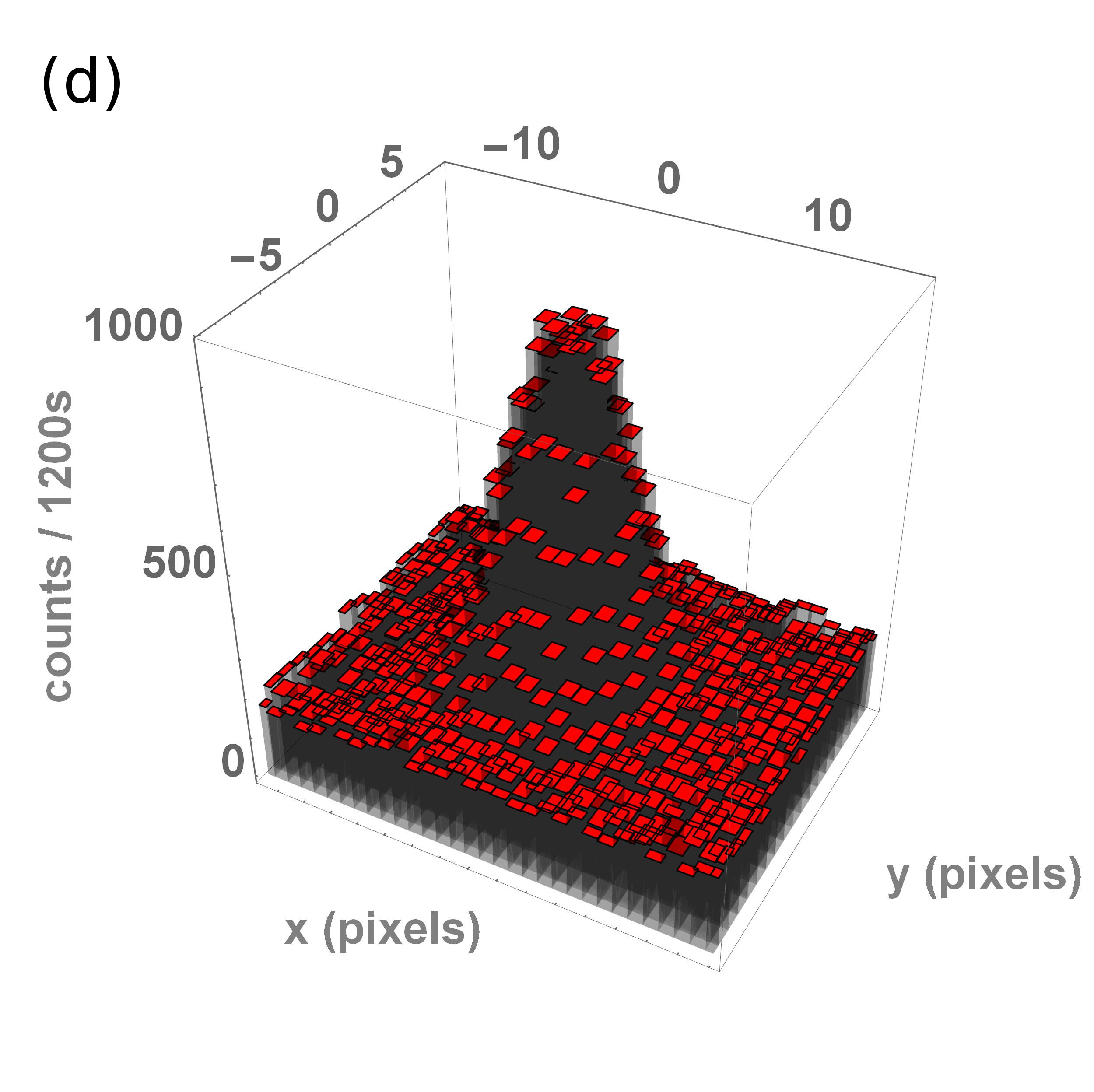}
    \end{subfigure}

    \begin{subfigure}[b]{0.48\textwidth}
        \centering
        \includegraphics[width=\textwidth]{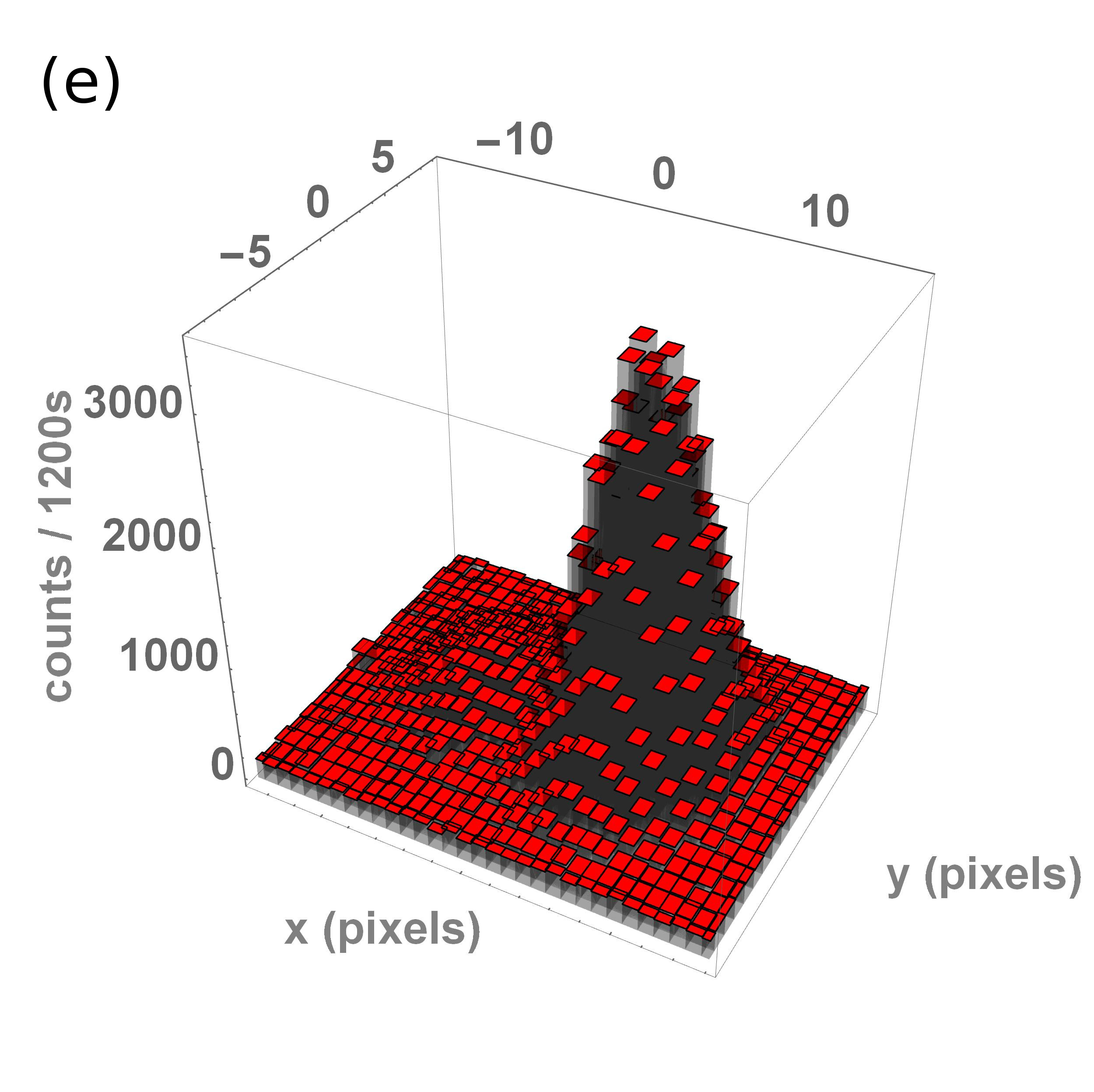}
    \end{subfigure}
    \begin{subfigure}[b]{0.48\textwidth}
        \centering
        \includegraphics[width=\textwidth]{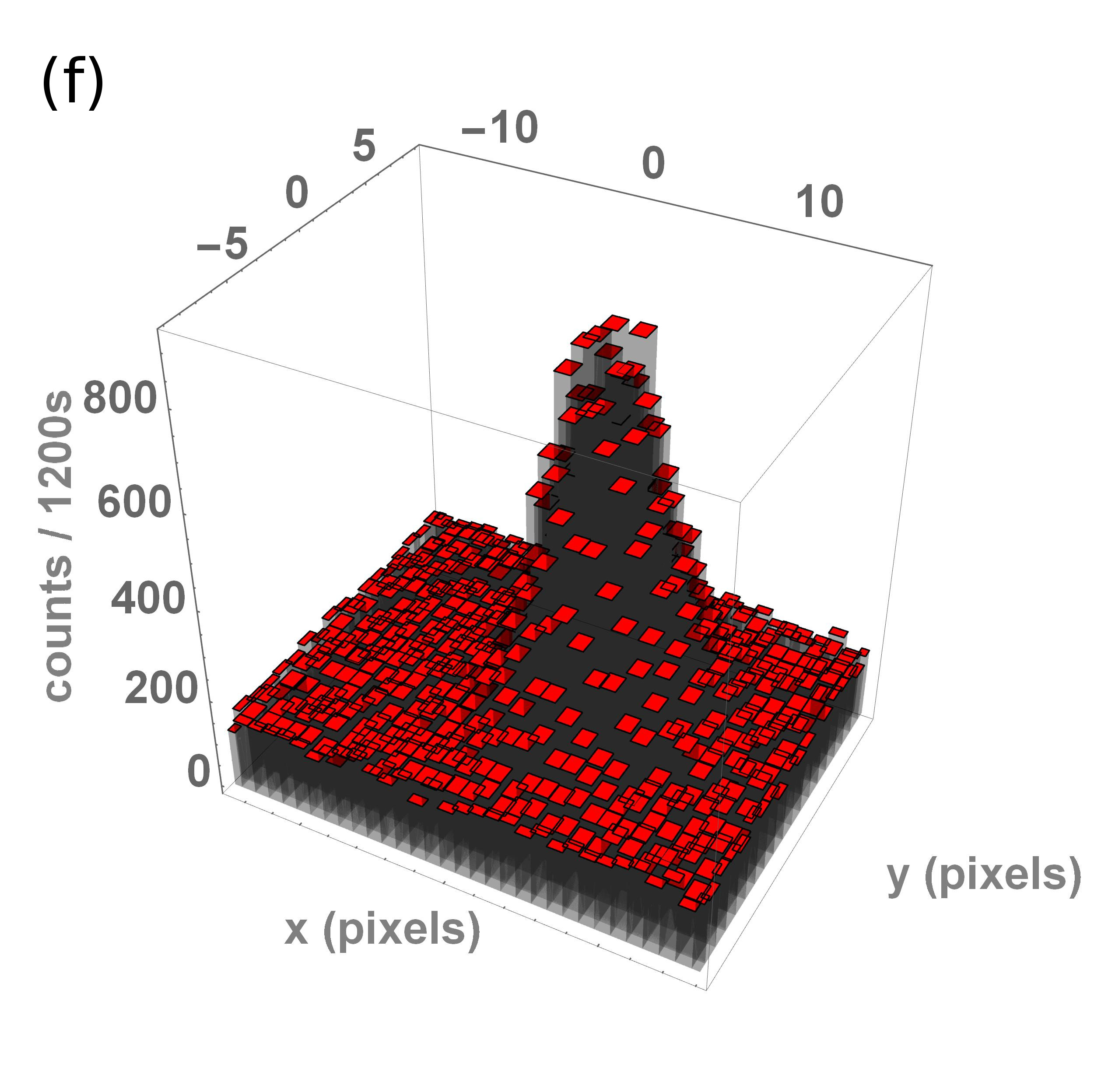}
    \end{subfigure}
	\caption{Plots of the photon counts distributions obtained with projective (\textbf{a},\textbf{c},\textbf{d}) and protective (\textbf{b},\textbf{d},\textbf{f}) measurements, for three different linearly-polarized initial states $\ket{\psi_\theta}$. \textbf{a}-\textbf{b}: ${\ket{\psi_{\frac{17}{60}\pi}} \approx 0.629\ket{H}+0.777\ket{V}}$; \textbf{c}-\textbf{d}: ${\ket{\psi_{\frac{\pi}{4}}} \approx 0.707\ket{H}+0.707\ket{V}}$; \textbf{e}-\textbf{f}: ${\ket{\psi_{\frac{\pi}{8}}} \approx 0.924\ket{H}+0.383\ket{V}}$.}
\label{distrib3D}
\end{figure}
Plots \ref{distrib3D}a, \ref{distrib3D}c, \ref{distrib3D}e show the results obtained for the three states in the PJ case.
We can see that the photons accumulate around the two eigenvalues positions $x_H$ and $x_V$, hence we can statistically find the expectation value as the counts ratio in Eq. (8).\\

Plots \ref{distrib3D}b, \ref{distrib3D}d, \ref{distrib3D}f host, instead, the PM results, in which all the photons accumulate in a specific position corresponding to $x=a\ev*{A}$, in agreement with our expectations.
This is clear evidence that, with protective measurements, each single photon carries information about the expectation value of its polarization.\\

\begin{figure}[!h]
\begin{subfigure}[b]{0.48\textwidth}
	\centering
	\includegraphics[width=\textwidth]{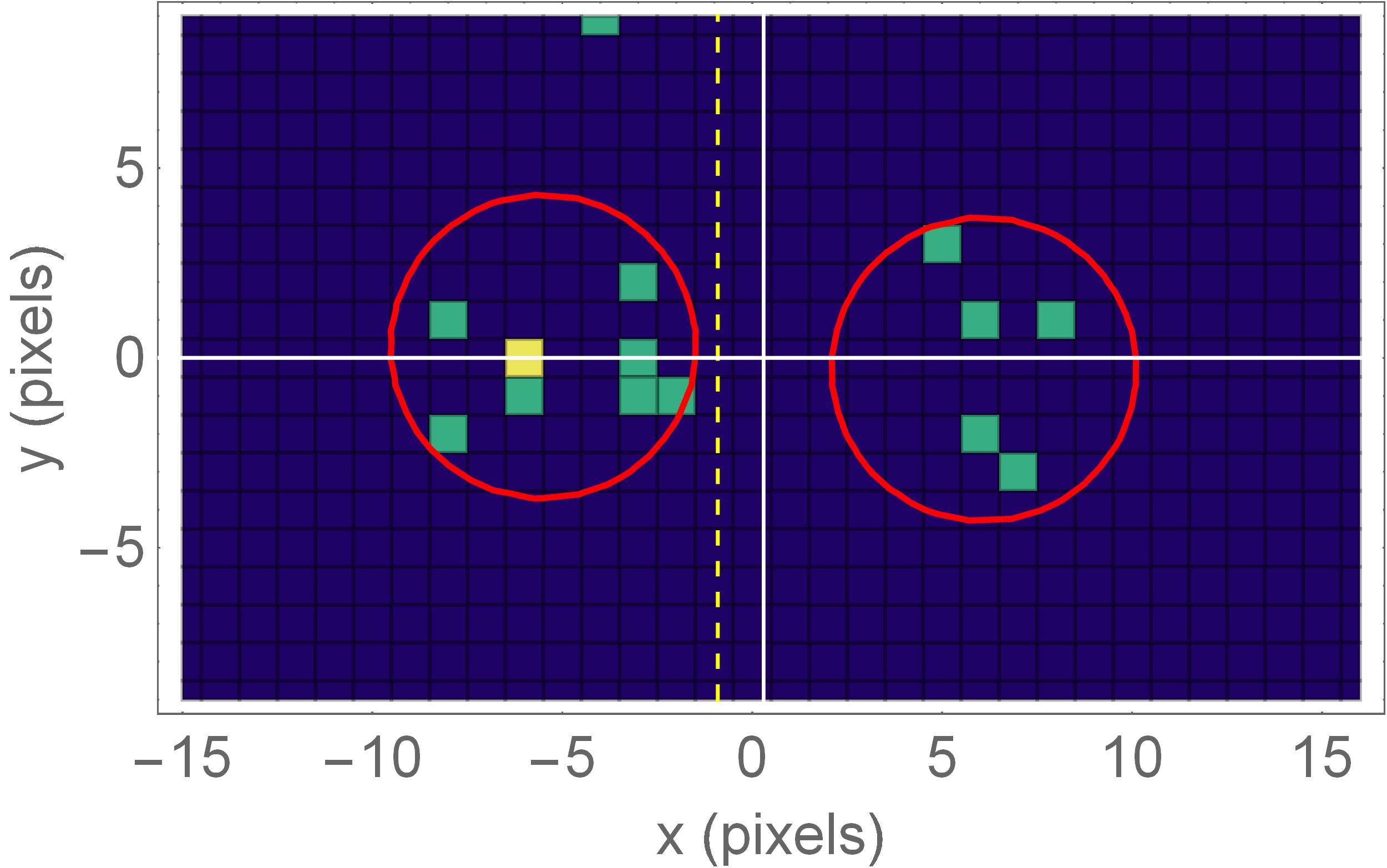}
	\caption{Projective measurement case (14 detection events).}
\end{subfigure}
\begin{subfigure}[b]{0.48\textwidth}
	\centering
	\includegraphics[width=\textwidth]{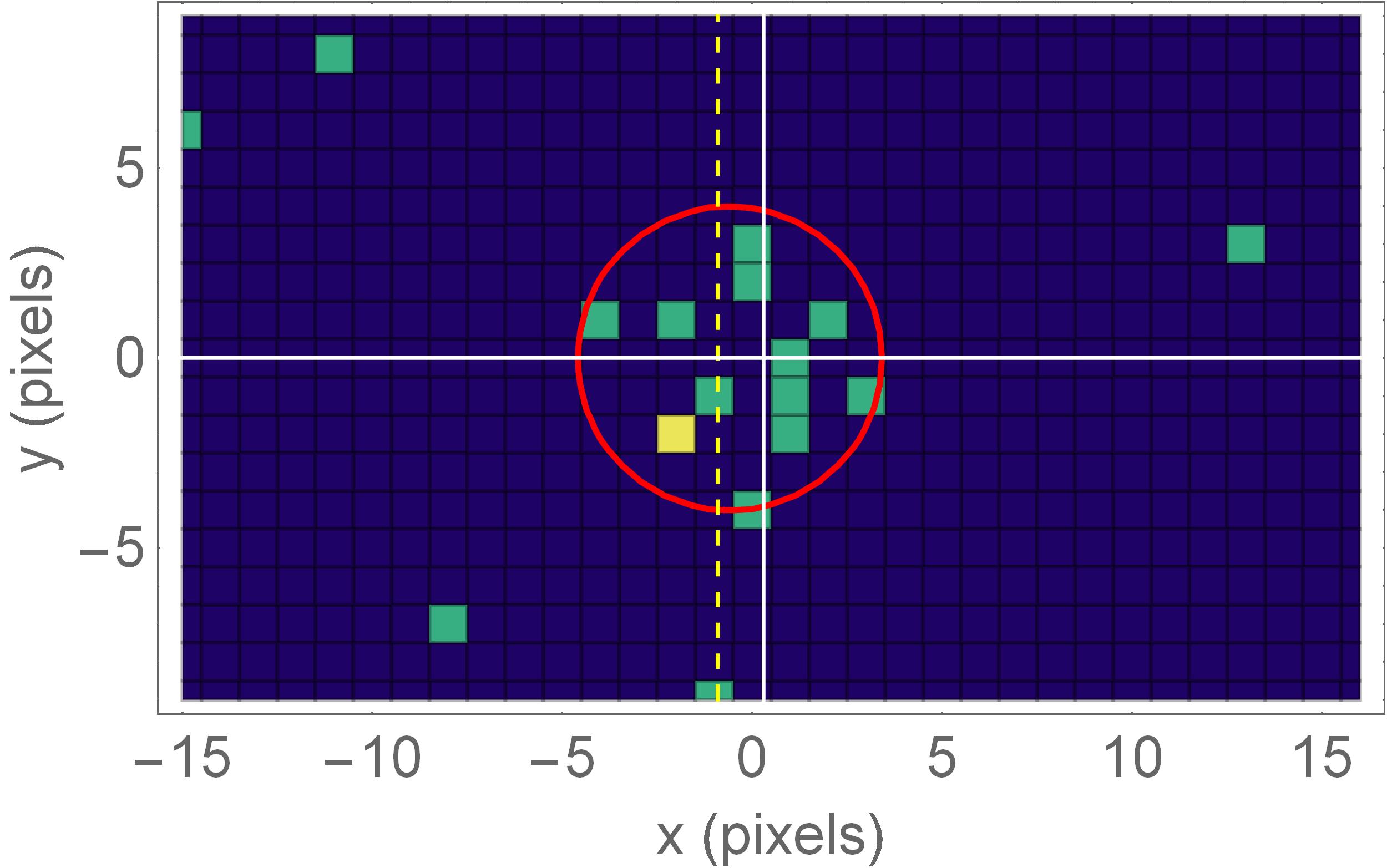}
	\caption{Protective measurement (17 detection events).}
\end{subfigure}
\caption{Few-event photon counts distributions for the input state $\ket{\frac{17}{60}\pi}$, with the first detection event is marked in yellow. Yellow dashed line: $x$ position corresponding to the theoretical expectation value of the polarization $\ev*{A}^{\mathrm{th}}=-0.208$. Red circles: FWHM of the corresponding distributions for the state $\ket{\frac{17}{60}\pi}$ reported in Fig. \ref{distrib3D}.}
\label{distribFew}
\end{figure}

The extracted expectation values are reported in Table \ref{expA}, column 3 for PJ and in Table \ref{expA}, column 4 for PM.
A detailed description of the expectation values and uncertainties analysis can be found in Appendix A.
We notice that the expectation values extracted with the PM method are in excellent agreement with the ones obtained with the traditional (PJ) one, as well as with the theoretically-expected values.\\
\begin{table}[h]
\caption{Comparison between experimental and theoretical expectation values. Experimental expectation values. $\ev*{A}^{\mathrm{th}}$: theoretical expectation value. $\ev*{A}^{\mathrm{PJ}}$: experimental expectation value with projective measurements. $\ev*{A}^{\mathrm{PM}}$: experimental expectation value with protective measurements.}
\centering
\begin{tabular}{cccc}
\toprule
\textbf{State}	& $\ev*{A}^{\mathrm{th}}$ & $\ev*{A}^{\mathrm{PJ}}$ & $\ev*{A}^{\mathrm{PM}}$\\
\midrule
$\ket{+}$ & 0 & -0.03(4) & 0.0012(14) \\
$\ket{\frac{17}{60}\pi}$ & -0.208 & -0.21(2) & -0.19(2) \\
$\ket{\frac{\pi}{8}}$ & 0.707 & 0.72(2) & 0.72(2) \\
\bottomrule
\end{tabular}
\label{expA}
\end{table}
To further confirm this result, the two plots in Fig. \ref{distribFew} show the equivalent of the distributions in Figs. \ref{distrib3D}a and \ref{distrib3D}b for few detection events.
While, for the PM in Fig. \ref{distribFew}b, all the counts (except for dark counts) belong to a region centered in the $x$ axis position corresponding to the expectation value $\ev*{A}$, the same does not happen for the PJ in Fig. \ref{distribFew}a.
Henceforth, with PM one can achieve a sound estimate of the average polarization value for the state $\ket{\frac{17}{60}\pi}$ from just the first detection event (yellow pixel in Fig. \ref{distribFew}b), obtaining $\ev*{A}=-0.3\pm0.3$, where the uncertainty is estimated from the width of the initial (Gaussian) spatial photon distribution.
This result is in agreement with the theoretically-expected value $\ev*{A}^{\mathrm{th}}=-0.208$.
On the contrary, it is not possible to do the same in the projective measurement case, since the first detection event (yellow pixel in Fig. \ref{distribFew}a) does not provide any reliable information about the expectation value of the detected photon polarization.
This is a final demonstration of the PM capability of extracting the expectation value even from a single detection event.
\section{Conclusions}
Protective measurements represent a novel, groundbreaking measurement paradigm, which allows preserving the initial coherence of the measured state and  extracting the expectation value of an observable, so far considered as a purely statistical quantity, even from a single detection event.
The presented results demonstrate this unprecedented capability by exploiting certified single photons.
In particular, in this paper we have described in detail the first experimental implementation of PM, providing the readers with all the details needed for a full understanding of the experiment and obtained results, as well as of the related implications for quantum mechanics foundations.
We verified that PM preserves the coherence of the initial state during the whole measurement process, as certified by the high Fidelities between the initially-prepared states and the reconstructed ones outgoing the PM.
Furthermore, the ability of PM to extract the expectation value of a quantum observable from a single (protected) particle is demonstrated by the photon distributions obtained with this protocol, always centered in a position proportional to the expectation value of the polarization of the detected photons and allowing to estimate polarization expectation values always in agreement with the ones obtained with traditional projective measurements, all matching the theoretical expectations within the experimental uncertainties.\\
These results shed important insight on the very foundations of quantum mechanics, especially in the long-standing debate about the ontic or epistemic nature of the wavefunction, at the same time paving the way toward new quantum measurement methods with possible significant application to quantum technologies and, in particular, to quantum metrology\cite{giovannetti,genovese,polino}, with the eventual possibility to exceed the quantum Cram\'{e}r-Rao bound \cite{parisCRB} thanks to parameter dependence of the measurement procedure \cite{seveso1,seveso2} .

\vspace{6pt}



\section*{Author contributions}
Conceptualization, E.C., I.P.D., L.V., M.Gen. and M.Gram.; methodology, A.A., E.C., F.P., I.P.D., L.V., M.Gen. and M.Gram.; software, A.A., E.R., F.P., I.P.D., R.L.; validation, I.P.D. and M.Gram.; formal analysis, E.R.; investigation, A.A., E.R. F.P.; resources, A.T., F.V., R.L.; data curation, F.P.; writing--original draft preparation, all authors; writing--review and editing, all authors; visualization, A.A., E.R. and F.P.; supervision, I.P.D., G.B., M.Gram. and M.Gen.; project administration, L.V. and M.Gen.; funding acquisition, E.C., I.P.D., M.Gen. and L.V. All authors have read and agreed to the published version of the manuscript.

\section*{Acknowledgments}
This work has been financed by the projects 17FUN01 `BeCOMe' and 17FUN06 `SIQUST', both funded from the EMPIR programme co-financed by the Participating States and from the European Union's Horizon 2020 research and innovation program.

\section*{Appendix: expectation value analysis}
Here we describe more in detail our analysis of the expectation values in this work.
The first step consists of extracting the centres $x_H$ and $x_V$ of the photon distributions corresponding to the horizontally- and vertically-polarized photons (i.e. the points corresponding to the $A$ eigenvalues $\pm1$), respectively.
This can be done by performing a linear regression of the acquisition for the states $\ket{H}$ and $\ket{V}$ and averaging over multiple acquisitions.\\

From the extracted $x_H$ and $x_V$, we define the center of our ``laboratory system'' (i.e. the point at $\ev*{A}=0$) as:
\begin{equation}
  x_0 = \frac{x_H+x_V}{2}
  \label{eq:x0-protective}
\end{equation}
and the distance between the center and one of the two extremes as:
\begin{equation}
  a=\frac{x_H - x_V}{2}
  \label{eq:a-protective}
\end{equation}
with an associated uncertainty of
\begin{equation}
  \sigma_a = \sigma_{x_0} = \frac{\sqrt{\sigma^2_{x_H}+\sigma^2_{x_V}}}{2}
  \label{eq:incertezza-x0-a}
\end{equation}

This allows evaluating the expectation values from both measurement procedures.
\subsection{Projective measurements}
In the projective measurement case, we find the expectation value as the counts ratio:
\begin{equation}
  \ev*{A}=\frac{(N_{H}-N_{H}^{(dark)})-(N_{V}-N_{V}^{(dark)})}{N_{H}+N_{V}-N_{H}^{(dark)}-N_{V}^{(dark)}}
  \label{eq:ev-stat}
\end{equation}
where $N_{\mathrm{V(H)}}$ is the number of counts in the region corresponding to the vertical (horizontal) polarization component and $N_{\mathrm{V(H)}}^{(dark)}$ is the number of dark and background counts in the same region, estimated by evaluating the number of counts outside the region of interest of the detector.

The associated uncertainty is
\begin{equation}
  \sigma_{\ev*{A}} = \sqrt{ \sum_{k=1}^{4} \left( \pdv{A}{N_k} \right)^2 \sigma_{N_k}^2 }
  \label{eq:ev-trad-unc}
\end{equation}
with $N_1=N_{V}$, $N_2=N_{H}$, $N_3=N_{V}^{(dark)}$, $N_4=N_{H}^{(dark)}$.
We evaluate the uncertainties on the number of background counts $N_{\mathrm{V(H)}}^{(dark)}$ by assuming a poissonian behaviour (i.e. $\sigma_{N_{V(H)}^{(dark)}} = \sqrt{N_{V(H)}^{(dark)}}$).
The uncertainty on the number of photons in the two regions, instead, is more delicate, as the two distributions belonging to horizontally- and vertically-polarized photons are separated, but not completely. For this reason, we evaluate these uncertainties as
\begin{align}
  \sigma_{N_{H}} &= \sqrt{N_{H} + (c_{H} \  N_{H})^2 + (c_{V} \  N_{V})^2}\\
  \sigma_{N_{V}} &= \sqrt{N_{V} + (c_{H} \  N_{H})^2 + (c_{V} \  N_{V})^2}
\end{align}
where the two coefficients $c_{\mathrm{H}}$ and $x_{\mathrm{V}}$ come from an ad hoc evaluation of the influence of the distribution tails (small, but still relevant) on the number of photon counts.

\subsection{Protective measurements}
In the protective measurement case, each photon carries information about the expectation value, estimated as the ratio:
\begin{equation}
  \ev*{A}= \frac{x-x_0}{a}
  \label{eq:ev-prot}
\end{equation}
being $x$ the position of the photon, corrected by compensating for unwanted deviations induced by the polarizers and $a=g/2$.
We extract $x$ from every pixel and then average it, weighting on the number of counts for each pixel.
The associated uncertainty is
\begin{equation}
\sigma_{\ev*{A}} = \Bigg[ \sigma_{\mathrm{ave}} + \left( \frac{1}{x_H^\prime-x_V^\prime} + \frac{\ev*{\hat{A}}}{x_H^\prime-x_V^\prime} \right)^2 \sigma_{x_H^\prime}^2 + \left( \frac{1}{x_H^\prime-x_V^\prime} - \frac{\ev*{\hat{A}}}{x_H^\prime-x_V^\prime} \right)^2 \sigma_{x_V^\prime}^2  \Bigg]^{1/2}
\end{equation}
where $\sigma_{\mathrm{ave}}$ indicates the standard deviation of the mean of $\ev*{A}$.
The second and third terms are the uncertainties on the parameters $x_{H(V)}^\prime = x_{H(V)} +x_{pol} -x_{void}$, where $x_{void}$ and $x_{pol}$ are the positions of the beam in the acquisition with a free optical path and with only the polarisers inserted, respectively.
This allows us to compensate for the aforementioned unwanted polarisers-induced deviations.
The variances associated with these parameters are $\sigma_{x_{H(V)}^\prime}^2 = \sigma_{x_{H(V)}}^2 + \sigma_{x_{pol}}^2 + \sigma_{x_{void}}^2$.\\




\end{document}